# Blueprint for a Rotating Universe

G. Chapline
Lawrence Livermore National Laboratory, Livermore, CA 94550

**Abstract**
A solution of the vacuum Einstein equations with a cosmological constant is exhibited which can perhaps be used to describe the interior of compact rotating objects, and may also provide a description of our universe on length scales approaching the size of the de Sitter horizon.

In this note we offer a solution to two outstanding problems in theoretical astrophysics. The first problem is to describe the structure of space-time inside a rotating object that is sufficiently compact that it lies entirely inside a surface where classical general relativity predicts that an event horizon would form. The conventional view is that such objects are "black holes". However, it has long been felt that the interior Kerr solution for compact objects is unphysical. More recently it has been argued [1,2] that the existence of black holes in general is inconsistent with quantum mechanics, and that in reality the black hole event horizon is actually a physical surface where classical general relativity fails and observable physical effects take hold [3]. Space-time outside this surface would in the case of a non-rotating object be described by the usual exterior Schwarzschild solution, while in references 1 and 2 it was proposed that the interior space-time could at least approximately be described using de Sitter's "interior" cosmological solution [4]. This guess is in accord with the expectation [5] that in order to be consistent with quantum mechanics the interior space-time of a collapsed object should be obtained by continuous "squeezing" of a condensate vacuum state. On macroscopic length scales a squeezed vacuum should locally resemble classical flat space-time with a non-vanishing vacuum energy; i.e. a region of de Sitter space-time.

One nagging question concerning the proposals of references 1 and 2, though, has been the question as to what should replace de Sitter's interior solution in the case of a rotating compact object. In the absence of such an interior solution it cannot be said that the idea of a star supported by vacuum energy matches the mathematical elegance of the black hole hypothesis in providing a description for both rotating and non-rotating compact objects.

In the following we address this deficiency by suggesting a mathematically simple candidate metric for the interior of a compact rotating object.

This metric provides a possible solution to another cosmological puzzle; namely, what would our universe look like on very large length scales if it were rotating? As it happens there are no solutions of the classical Einstein equations that contain rotating matter and are consistent with quantum mechanics, because generically there is no definition of a universal time in rotating classical space-times. Indeed, this was one of the main motivations for Godel in introducing his famous model for a rotating universe [6]. In addition to the problem of universal time, rotating space-times such as the Godel universe also typically contain closed time-like curves, which creates problems with causality.

Recently it has been suggested [7] that the way to resolve the difficulties with classical rotating space-times is to suppose that the rotation is actually carried by spinning space-time strings [8], in a manner analogous to the way rotation of superfluid helium inside a rotating container is carried by quantized vortices. The spinning strings resolve the question of the consistency of rotating space-times with quantum mechanics because the vorticity of space-time would be concentrated into the cores of the spinning strings where general relativity fails. It can be shown that averaging over the vorticity of many perfectly aligned spinning strings leads to a Godel-like space-time [7]. However, it was argued in ref. 7 that Godel-like cosmologies are actually not physically realizable, at least on cosmological length scales, because the spinning springs are massless and there is no natural force which would keep them perfectly aligned. This means that except possibly near to a boundary for the rotating space-time the spinning strings would tend to become tangled, and so Godel-like solutions of the classical Einstein equations are physically unstable. Of course, the astrophysical relevance of the Godel's cosmological solution is *prima facie* questionable because Godel's solution is both cylindrically symmetric and stationary. The observable universe is neither stationary nor cylindrically symmetric. On the other hand, there is observational evidence that the vacuum energy in fact does not vary with time. This means that the vacuum energy can be modeled as a cosmological constant, and at least asymptotically there is a stationary event horizon. Therefore it may be quite relevant to ask what space-time should look like in a stationary rotating universe on cosmological length scales. In the following we offer an possible answer to this question.

In both the cosmological case and for the case of a rotating compact objects that lie inside their event horizon, what we are seeking is a rotating analog of de Sitter's "interior" solution [4] for a space-time with no matter but a cosmological constant $\Lambda$. Our proposed interior metric is (we use units such that $8\pi G/c^2 = 1$)

$$ds^2 = [1 - \frac{\Lambda}{3}(r^2 - a^2\cos^2\theta)]dt^2 + 2a(1 - \frac{\Lambda}{3}r^2\cos^2\theta)dtd\varphi$$
$$- \frac{r^2 + a^2\cos^2\theta}{r^2 - \frac{\Lambda}{3}r^4 + a^2}dr^2 - \frac{r^2 + a^2\cos^2\theta}{1 - \frac{\Lambda}{3}a^2\cos^2\theta\cot^2\theta}d\theta^2 \quad (1)$$
$$-(r^2\sin^2\theta - a^2\cos^2\theta)d\varphi^2,$$

where $a$ is the angular momentum per unit mass. This metric is a limiting case of a class of metrics discovered by Carter [9]. In the limit $a \to 0$ the metric (1) reduces to de Sitter's 1917 metric. It is clear by inspection that in contrast with the interior Kerr metric there are no space-time singularities near to $r = 0$ for any value of $\theta$. The apparent singularities in the $g_{rr}$ and $g_{\theta\theta}$ components of the metric tensor can be removed by a change of variables [10], and represent event horizons where $g_{00}g_{\varphi\varphi} - g_{0\varphi}^2 = 0$. The singularity in $g_{rr}$ is associated with a spherical event horizon located at

$$r_H^2 = \frac{3}{2\Lambda} + \left[\frac{9}{4\Lambda^2} + \frac{3a^2}{\Lambda}\right]^{1/2}. \quad (2)$$

In the limit $a \to 0$ $r_H$ becomes the de Sitter horizon $\sqrt{3/\Lambda}$. In addition to the spherical event horizon (2) there is a conical event horizon located at

$$\tan^2\theta_H = -\frac{1}{2} + \sqrt{\frac{\Lambda}{3}a^2 + \frac{1}{4}}. \quad (3)$$

In the case of slow rotation $\Lambda a^2 \ll 1$ this conical event horizon is located very near to the axis of rotation.

The nemesis of rotating space-times, closed time-like curves, will appear if $g_{\varphi\varphi}$ is positive for some values of $r$. In our case this means that

$$r^2\sin^2\theta - a^2\cos^2\theta < 0. \quad (4)$$

This is will be satisfied if $\rho<a$, where $\rho$ is the horizontal distance from the axis of rotation. Thus for slow rotation closed time-like curves will appear very close to the rotation axis. This is very reminiscent of the situation with spinning strings [8]. Actually for all values of the rotation parameter $a$ the conical horizon (3) lies inside the region where closed time-like curves appear, and the critical angle where the inequality in (4) becomes an equality is precisely the horizon angle $\theta_H$ when $r = r_H$. Thus the conical horizon delineates a polar region of the spherical horizon where classical general relativity fails.

If we think of the metric (1) as the "interior" space-time for a compact rotating object, then it would be reasonable to suppose that the "exterior" space-time outside the horizon (2) is just the Kerr solution. The event horizon for the Kerr solution will match the event horizon (2) if the mass parameter for the Kerr solution is

$$m = \frac{\Lambda}{6} r_H^3. \tag{5}$$

Curiously this is the same condition that was used in ref. 1 to match de Sitter's interior solution to the exterior Schwarzschild solution in the case of a non-rotating compact object. Near the event horizon (2) the angular part of (1) has the form

$$ds^2 = -a^2 \frac{\sin^2\theta - \frac{\Lambda}{3} a^2 \cos^4\theta}{r_H^2 + a^2 \cos^2\theta} (dt - \frac{r_H^2}{a} d\phi)^2 - \frac{r_H^2 + a^2 \cos^2\theta}{1 - \frac{\Lambda}{3} a^2 \cos^2\theta \cot^2\theta} d\theta^2. \tag{6}$$

This metric is not exactly equal to the Kerr metric evaluated on the event horizon. However, with the matching condition (5) it is tolerably close to the Kerr metric on the event horizon for all values of $a$ such that $\Lambda a^2 < 1$, except near to the conical horizon where general relativity fails.

Inside the event horizon (2) the behavior of the metric (1) is completely different from the Kerr metric. As noted above, there are no space-time singularities. In the case of the Kerr solution $g_{00} < 0$. everywhere inside the "ergosphere" whose outer boundary lies at $r^2 + a^2 \cos^2\theta = 2mr$. Although our $g_{00}$ is negative at the event horizon ( and close to the Kerr $g_{00}$ ), it is positive for all values of $r$ inside the de Sitter horizon . Also in contrast with the Kerr solution our $g_{rr}$ is negative for all values of $r$ inside the spherical event horizon. Thus the problematic reversal of the roles of time and radial distance in the Kerr solution is alleviated.

In summary, the metric (1) is a plausible candidate for both the interior space-time of rotating compact objects and the large scale structure of a rotating universe. In particular, this metric provides a possible solution to the problems connected with the conflict between general relativity and quantum mechanics in compact rotating space-times. Our metric resembles the exterior Kerr metric near to the spherical event horizon, which in turn provides a new twist to the long-standing puzzle concerning the fate of matter in a steady state universe. The Kerr solution has the property that inside the ergosphere particles cannot be at rest but must rotate about the axis. At the event horizon the frame in which particles could be at rest rotates with the "frame dragging" angular velocity

$$\left.\frac{d\phi}{dt}\right|_{r=r_H} = \frac{a}{r_H^2 + a^2}. \qquad (7)$$

For the metric (1) $g_{00} < 0$ at the spherical event horizon and the frame rotation velocity is $a/r_H^2$, so particles in our space-time will also rotate as they approach the event horizon from the inside. Indeed our interior metric contains a reflection of the Kerr ergosphere whose inner boundary is at $r^2 = (3/\Lambda)^{1/2} + a^2 \cos^2\theta$. Reflection symmetry between the inner and outer metrics at an event horizon is a signature of replacing the mathematical event horizon with a quantum critical layer [1] with its attendant effects such as baryon decay [3]. In addition, particles entering an ergosphere can exchange energy and angular momentum with the space-time as a whole [11] – which in the cosmological case is the rotating universe. All of this makes the suggestion [12] that the observable expanding universe lies inside a rotating steady state universe even more interesting.

The author is very grateful for numerous discussions with Pawel Mazur.